\documentclass{aastex63}
\usepackage{lineno}
\usepackage{hyperref}
\usepackage[utf8]{inputenc}
\usepackage{amsmath}
\usepackage{amsfonts}
\usepackage{enumerate}
\usepackage{ytableau}
\usepackage{graphicx}% Include figure files
\usepackage{indentfirst}
\usepackage[english]{babel}
\usepackage{overpic}
\usepackage[normal,footnotesize]{caption2}
\def\baselinestretch{1.25}
\usepackage{appendix}
\usepackage{amssymb}
\usepackage{amsthm}
\usepackage{epsfig,graphicx,epstopdf}

\newcommand{\gray}{$\gamma$-ray\ }
\newcommand{\grays}{$\gamma$-rays\ }
\received{}
\revised{}
\accepted{}
\submitjournal{ApJL}

\shorttitle{Cas A upper limit}
\shortauthors{LHAASO Collaboration}
\begin{document}

\title{Does or did the supernova remnant  Cassiopeia A operate as a PeVatron?
}
%\linenumbers
%\collaboration{The LHAASO Collaboration }
\author{Zhen Cao}
\affiliation{Key Laboratory of Particle Astrophysics \& Experimental Physics Division \& Computing Center, Institute of High Energy Physics, Chinese Academy of Sciences, 100049 Beijing, China}
\affiliation{University of Chinese Academy of Sciences, 100049 Beijing, China}
\affiliation{Tianfu Cosmic Ray Research Center, 610000 Chengdu, Sichuan,  China}
 
\author{F. Aharonian}
\affiliation{Dublin Institute for Advanced Studies, 31 Fitzwilliam Place, 2 Dublin, Ireland }
\affiliation{Max-Planck-Institut for Nuclear Physics, P.O. Box 103980, 69029  Heidelberg, Germany}
 
\author{Q. An}
\affiliation{State Key Laboratory of Particle Detection and Electronics, China}
\affiliation{University of Science and Technology of China, 230026 Hefei, Anhui, China}
 
\author{Axikegu}
\affiliation{School of Physical Science and Technology \&  School of Information Science and Technology, Southwest Jiaotong University, 610031 Chengdu, Sichuan, China}
 
\author{Y.X. Bai}
\affiliation{Key Laboratory of Particle Astrophysics \& Experimental Physics Division \& Computing Center, Institute of High Energy Physics, Chinese Academy of Sciences, 100049 Beijing, China}
\affiliation{Tianfu Cosmic Ray Research Center, 610000 Chengdu, Sichuan,  China}
 
\author{Y.W. Bao}
\affiliation{School of Astronomy and Space Science, Nanjing University, 210023 Nanjing, Jiangsu, China}
 
\author{D. Bastieri}
\affiliation{Center for Astrophysics, Guangzhou University, 510006 Guangzhou, Guangdong, China}
 
\author{X.J. Bi}
\affiliation{Key Laboratory of Particle Astrophysics \& Experimental Physics Division \& Computing Center, Institute of High Energy Physics, Chinese Academy of Sciences, 100049 Beijing, China}
\affiliation{University of Chinese Academy of Sciences, 100049 Beijing, China}
\affiliation{Tianfu Cosmic Ray Research Center, 610000 Chengdu, Sichuan,  China}
 
\author{Y.J. Bi}
\affiliation{Key Laboratory of Particle Astrophysics \& Experimental Physics Division \& Computing Center, Institute of High Energy Physics, Chinese Academy of Sciences, 100049 Beijing, China}
\affiliation{Tianfu Cosmic Ray Research Center, 610000 Chengdu, Sichuan,  China}
 
\author{J.T. Cai}
\affiliation{Center for Astrophysics, Guangzhou University, 510006 Guangzhou, Guangdong, China}
 
% ...




\begin{abstract}
For decades,  supernova remnants (SNRs) have been considered the prime sources of Galactic Cosmic rays (CRs). But whether SNRs can accelerate CR protons to PeV energies and thus dominate CR flux up to the {\it knee} is currently under intensive theoretical and phenomenological debate. The direct test of the ability of SNRs to operate as CR PeVatrons can be provided by ultrahigh-energy  (UHE; $E_\gamma \geq 100$~TeV) $\gamma$-rays. In this context, the historical SNR Cassiopeia A (Cas A) is considered  one of the most promising target for UHE observations. This paper presents the observation of Cas A and its vicinity by the LHAASO KM2A detector. The exceptional sensitivity of LHAASO KM2A in the UHE band, combined with the young age of Cas A, enabled us to derive stringent model-independent limits on the energy budget of UHE protons and nuclei accelerated by Cas A at any epoch after the explosion. The results challenge the prevailing paradigm that Cas A-type SNRs are major suppliers of PeV CRs in the Milky Way.

\end{abstract}



\section{Introduction} 
Within the current Galactic CR paradigm, SNRs are considered the major contributors to the observed CR flux. For decades, this conviction has been  based on phenomenological arguments and supported by the Diffusive Shock Acceleration (DSA) theory (for a review, see \cite{MalkovDrury}). The dominance of SNRs' contribution up to the "knee", a distinct break in the CR spectrum around 3-4~PeV, would imply that at certain epochs of their evolution, SNRs should operate as PeVatrons. However, as has been recognised long ago \citep{LagageCesarsky}, within the framework of the standard DSA applied to young SNRs, the energy of accelerated protons  could hardly exceed 100~TeV. A possible solution was proposed by \cite{Bell2004} by amplifying the magnetic field upstream of the shock through the instabilities driven by CRs. Combined with two other critical conditions, (1) shock speed of thousands of km/s, and (2) particle diffusion in the extreme (Bohm) regime, enhancing the magnetic field  to $B \geq 100 \mu \rm G$ allows acceleration of protons and nuclei to 1~PeV/nuc.

The interactions of the shock accelerated protons and nuclei with the ambient gas make 
young SNRs potentially detectable sources of $\gamma$-rays and neutrinos. Therefore, the detection of TeV $\gamma$-rays from more than a dozen young SNRs \footnote{see the online catalogue for TeV $\gamma$-ray sources,  http://tevcat.uchicago.edu/}  is a direct proof of effective acceleration of protons and/or electrons to very high energies.  
In general, the available data do not allow us to distinguish between the hadronic (``$\pi^0$-decay") and leptonic (``inverse Compton") origin of TeV $\gamma$-rays. Even ignoring this ambiguity, i.e., assuming that $\gamma$-rays are produced by accelerated protons,  we face a problem with the explanation of TeV $\gamma$-ray measurements. The reported steep  $\gamma$-ray spectra with a differential power-law index $\Gamma \sim 2.4-2.7$ \citep[e.g., as summerised in the supplementary of][]{aharonian19} do not readily agree with predictions of the standard DSA theory for strong shocks of young SNRs. 
This result can be interpreted in three different ways:

(1) A steep power-law proton spectrum is mimicked by the result of the combination of a hard ($E^{-2}$ type) power-law spectrum and an "early" exponential cutoff, $E_0 \ll 100$~TeV.  This simple explanation supports the predictions of the standard DSA and the conclusion of \citet{LagageCesarsky} for the case of a non-amplified magnetic field. This mathematical trick works in a relatively narrow (one decade or so) energy interval around and below the cutoff energy. If this explanation is correct, at higher energies the measurements should reveal a gradual spectral steepening in the deep exponential cutoff region. 

(2) The claim that strong shock acceleration results in a hard acceleration spectrum is not always correct.  Indeed, steep power-law proton spectra can be formed over a wide energy interval in certain realistic environments and scenarios  \citep[see, e.g.,][]{bell11,Bell_steep,Malkov_steep,Malkov_steep2,Caprioli_steep,xu22}. The steep power-law spectra can extend over a wide energy interval, formally up to 1 PeV, assuming that such energetic protons are effectively confined in the shell. Correspondingly, the resulting steep power-law $\gamma$-ray spectra could continue up to 100~TeV. 
The detection of steep multi-TeV $\gamma$-ray spectra in both above scenarios is a rather difficult but feasible task for detectors like CTA and LHAASO.

(3) Acceleration of protons to PeV energies at the very early ($\ll 100$~yr) epoch of the SNR evolution is preferable \citep{bell13,zirak14}, in particular, because of the shock speed exceeding 10,000 km/s  and the high density of the progenitor wind.  However, the CRs with highest energy have already escaped from the shock upstream and thus the remnant is currently emptied of multi-TeV CRs.   This can naturally explain the steep spectra of $\gamma$-rays from SNRs, including from very young ones, in particular  from  SNR Cassiopeia A \citep{casa_hegra,casa_magic, 
casa_veritas} and Tycho \citep{tycho_veritas}. 

 The $\gamma$-ray spectrum in scenario (2) is predicted to have a steeper power law shape without cutoff, which is significantly different from the other two scenarios. The $\gamma$-ray spectra from the remnants in scenarios (1) and (3) could be similar,  although for different reasons. On the other hand, scenario (3) differs principally from scenario (1) by the $\gamma$-radiation outside the remnant. While in scenario (1), PeV particles are not produced at all, scenario (3) allows proton acceleration up to 1~PeV, but assumes that presently the highest energy particles already left the remnant and presently occupy regions beyond the shell.  Any outcome of  $\gamma$-ray measurements at tens to hundreds TeV performed with detectors of adequate sensitivity - either detection of positive signals or upper limits - would provide definite answers to whether specific SNRs could accelerate protons 1~PeV at any epoch of their evolution. 

The detection or upper limits from SNRs of different types and ages could finally establish whether SNRs, as a source population, can be responsible for the CR fluxes up to the "knee". 
The results would be relevant to both scenarios (2) and (3).

The historical SNR Cas A represents a key target for such studies. It has a well-defined age and distance. It also belongs to a class of core-collapse supernova that explodes in a dense red supergiant wind. Such systems are believed to be able to accelerate ions to the highest energy due to the efficient amplification of magnetic fields by CR streaming in dense environments. 
%And the CR energy budget injected by such systems also dominates since their higher explosion energy. 

In this paper, using LHAASO KM2A data, we derive  \gray flux upper limits towards Cas~A and test its ability to act as a PeVatron. In Sec.2, we discuss the status of previous gamma-ray observations of Cas~A and the theoretical challenges of acceleration of protons 
to ultrahigh energies at different epochs of its evolution. In Sec.3, we review the physical properties of Cas~A, including the gas distribution in proximity to the remnant. In Sec.4, we describe the procedure of extraction of flux upper limits based on the LHAASO KM2A data. In Sec.5, we discuss the astrophysical implications of the obtained results.

%LHAASO is a large hybrid EAS array being constructed at Haizi Mountain, Daocheng, Sichuan province, China. It is composed of three sub-arrays, including the one square-km array (KM2A), Water Cherenkov Detector Array (WCDA) and the wide-field air Cherenkov/fluorescence telescopes (WFCTA) array \citep{article}. Benefiting from the excellent backgound rejection power and large filed of view, LHAASO is very suitable for extended source observation. Even though only a half of KM2A has been operated since the end of 2019, the sensitivity is already better than what have been achieved by previous observations above tens of TeV, and the detailed study of the performances of KM2A via the observation on the Crab Nebula is presented in \citep{2020arXiv201006205A}. 

%In this paper we present the LHAASO KM2A observations on Cas A. The non-detection of this source has already put stringent limit on the total VHE CR accelerated by this source and challenge the consensus that SNRs are the main contributor of the Galactic CRs. The paper are organized as follows: in Sec.2 we describe the gas and CR content in Cas A region, in Sec.3 we analyzed the LHAASO KM2A data in this region and derive the upper limit of VHE $\gamma$-ray emissions. In Sec.4 we discuss the results and the possible implications. 

\section{Basic facts about Cas A}

%The origin of Cosmic Rays (CRs) is still a mystery. It is believed that at least up to 1~PeV ($10^{15}~\rm eV$)  the CRs have an Galactic origin. In this regard, supernova remanants (SNRs) are considered as the most promising candidates for CR accelerators in our Galaxy. \citet{agile_w28,fermi_pion} have revealed the pion-bump feature in the $\gamma$-ray spectrum towards the mid-age SNRs interacting with molecular clouds, which is regarded as the strong proof that SNRs do accelerate CR protons. However, the $\gamma$-ray spectrum show a significant break at about 10~ GeV, which indicates the break or cutoff at about 100~GeV in the spectrum of the parent CR protons. Consequently, such system cannot be responsible to CRs up to PeV. 

%One solution of such tension is that the shock speed of these shock  of these mid-age SNRs have already been significantly deaccelerated and the maximum energy of accelerated CRs ($E_{max}$) is much smaller than those "very young" SNRs with much faster shock speed. \citet{schure13} have suggested the SNRs in dense environment will induce lager escaping current and more efficient magnetic field amplification, thus higher $E_{max}$. Cassiopeia A (Cas A) is regarded as one of the best candidate, which should accelerate CRs up to PeV in the first decade after explosion. 
Cassiopeia A (Cas A, SNR G111.7-02.1)  is one of the youngest SNRs in our Galaxy.
This $\approx 340$~year old \citep{reed95} structure is the remnant of  a type IIb supernova \citep{kraus08} located at a distance of 3.4~kpc. It is one of the brightest galactic radio sources  with a shell structure of angular radius 2.5' corresponding to the physical size of 2.5 pc \citep{kassim95}. The synchrotron radiation extends from radio \citep{tuffs97} to hard X-rays of about $100~\rm keV$ \citep{casa_nustar}. 
%The X-ray observations indicate that the reverse shock should be invoked in the particle accelerations \citep{sato18}. 
Fermi LAT observations reveal a hint of a hadronic origin of the $\gamma$-ray emissions \citep{yuan13,zirak14}. TeV $\gamma$-ray emission from 
Cas~A has been discovered by the HEGRA IACT array \citep{casa_hegra}, 
and later confirmed by the  MAGIC \citep{casa_magic} and VERITAS \citep{casa_veritas} collaborations.  

The hadronic models of GeV-to-TeV radiation require  a very high, but still acceptable, acceleration efficiency of about 25\% \citep{zirak14}, while in the two-zone models the acceleration efficiency required can be lower \citep{liu22}.
The significant  steepening or a cutoff in the spectrum reported around  a few TeV by  \citet{casa_magic}
and \citet{Abeysekara_2020} implies a corresponding steepening in the spectrum of parent protons at energies of tens of TeV. This constrains but does not exclude the presence of PeV protons in the nebula or outside the shell. Due to the limited age of Cas A, even if PeV protons have been accelerated at the early ($\ll 100$~years) epoch of the SNR, they could not propagate too far from the SNR. Thus, for robust conclusions, one should probe both the SNR itself and the $\approx 100$~pc environment surrounding Cas~A, for $\gamma$-rays of energy exceeding 100 TeV.

%due to the limited age of Cas A of about 340 years, if the very high energy (VHE) CRs are accelerated in the first decade after the explosion, it cannot propagate too far from the SNR. Thus the very high energy $\gamma$-ray observations can set  limit on the VHE CR protons accelerated by Cas A.

\section{Gas and CRs in the vicinity of Cas A}
\begin{figure}
    \centering
     \includegraphics[width=0.4\linewidth]{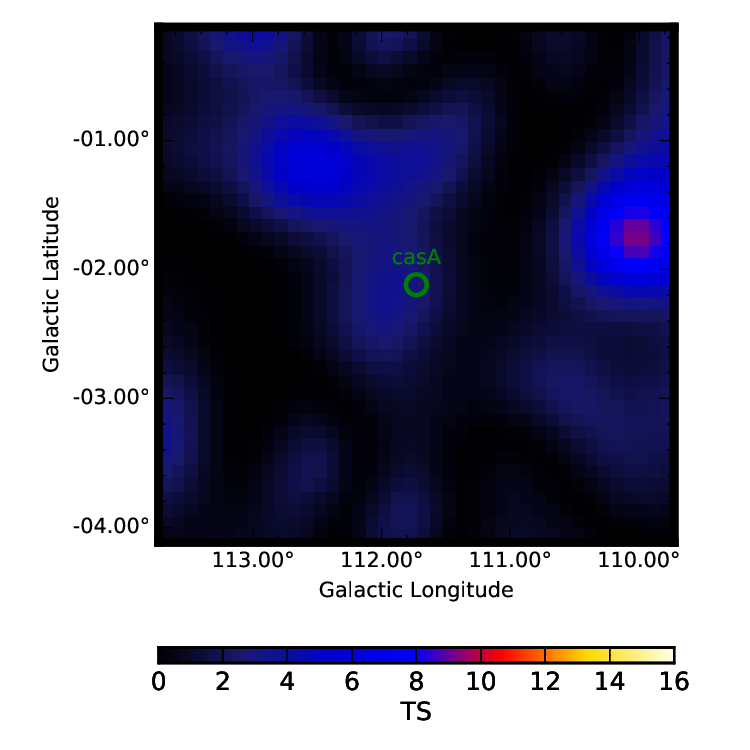}
    \includegraphics[width=0.4\linewidth]{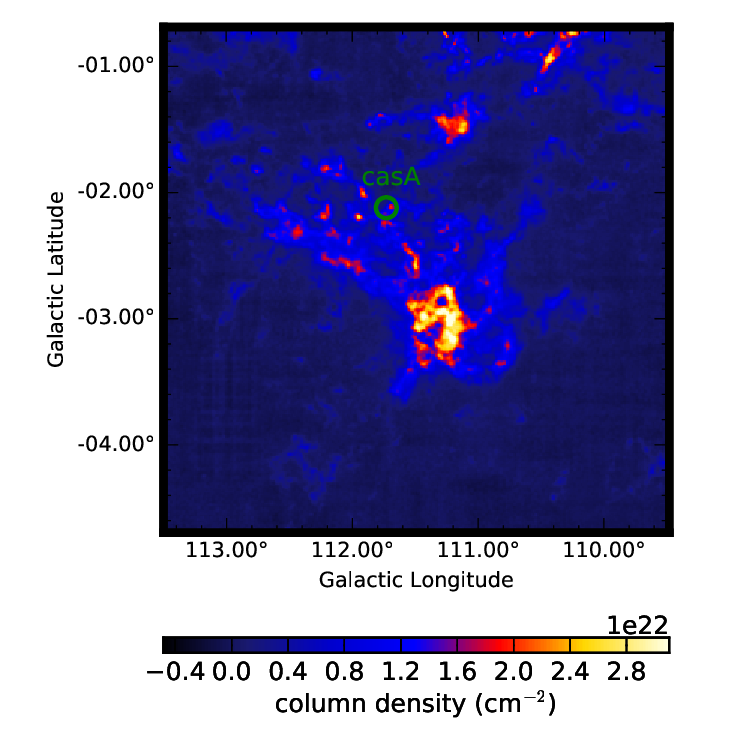}  
    \caption{{\it left panel}: The LHAASO-KM2A TS map above $25~\rm TeV$ in the $4^{\circ}\times4^{\circ} $ region around Cas A, the colorbar shows the TS value. 
    {\it right panel}:The gas column density near Cas A derived from $^{12}CO$ line observations \citep{ma19}. The green circle labels the position of Cas A.   }
    \label{fig:co}
\end{figure}

 Since we are searching for \grays  produced by PeV CR protons through interactions with the ambient gas, the information about the gas is crucial for our studies.  
 Cas~A is located within a dense gas environment. 
 \citet{zhou18} have detected molecular clouds  of  total mass $200~\rm M_{\rm \odot}$  in front of Cas A. 
 %The number density of these foreground molecular clouds is estimated as $1000~\rm cm^{-3}$. 
Within 200 pc around Cas~A, \citet{casa_ism} has derived the average gas density of about $3~\rm cm^{-3}$ by investigating the infrared emission. In this region, compact molecular clumps with a density as high as $10^5~\rm cm^{-3}$ have also been detected \citep{casa_mc}. \citet{ma19} have performed a detailed J = 1-0 survey of CO in a large field near Cas A  using the Purple Mountain Observatory (PMO) 13.7 m 
millimetre telescope. The total mass of molecular gas was $\approx 9.5\times 10^5~\rm M_{\odot}$, which corresponds to an average density of $10~\rm cm^{-3}$ within 100 pc proximity of Cas A. The line profiles have asymmetric or broadened shapes, which indicates a possible interaction between the SNR and the molecular clouds. In this work, we used the  J = 1-0 $^{12}\rm CO$ data as described in \citet{ma19}. The derived molecular gas column densities are shown in Figure.\ref{fig:co} by assuming $2.0\times 10^{20}~\rm (K ~km ~s^{-1})^{-1}$ for the  $\rm CO-to-H_2$ conversion factor  \citep{bolatto13}.

As mentioned above,  the average gas density is estimated as $10~\rm cm^{-3}$ in the vicinity, so to estimate the gas mass we need to calculate the volume that is occupied by CRs.  The latter depends on the CR propagation in the vicinity of Cas~A.  Since we deal with relatively small spatial scales and ultra-high energies,  we cannot {\it a priori} assume that the propagation proceeds in the diffusive regime. One cannot exclude that close to the accelerator, highest energy CRs propagate ballistically near the source and enter the diffusive regime  when $r>D/c$, where $r$ is the distance from the source, $D$ is the diffusion coefficient and $c$ is the speed of light. The diffusion coefficient in the Interstellar Medium (ISM)  is estimated as $D(E)\sim 2\times 10^{28} (\frac{E}{1~\rm GeV})^{0.5} ~\rm cm^2/s$ \citep{yuan17}.  Its extrapolation to $E = 1~\rm PeV$ gives $D \sim 2 \times 10^{31} ~\rm cm^2/s $.  This implies  $r \gtrsim D/c \simeq$ 200~pc. For comparison,  rectilinear propagation length of  protons  over 340 years is $\simeq$ 110 pc. Thus, even in the case of acceleration of PeV protons at the early epochs of the SNR, they would still continue to move ballistically.   
Due to the beaming effects, only the  protons propagating towards us  can produce visible $\gamma$-rays.  Then the \gray source should have a similar {\it angular}  size as the accelerator, i.e., the SNR shell. This is about   $2.5'$, so it can be detected by 
LHAASO-KM2A  as a point-like source. 

%Note that in this case, only the foreground gas with  a total mass  $\approx 200~\rm M_{\rm \odot}$ and  density as high as $1000~\rm cm^{-3}$ \citep{zhou18}, contributes to the detected $\gamma$-ray emission. 

The diffusion of CRs in the vicinity of the accelerators could be strongly suppressed due to  enhanced turbulence introduced, e.g.,  by the streaming instability of injected CRs (see e.g. \citep{malkov13}). The interaction of the cosmic ray precursor and dense clumps can also drive turbulence. By either following the turbulent tangling field lines or mirror diffusion in compressible magnetic fluctuations \citep{lazarian21}, high-energy CR protons might undergo very slow diffusion. In this case, PeV CRs could propagate in the diffusive regime and the region occupied by VHE CRs would be of the order $r \sim \sqrt{2DT } \sim 20 ~\rm pc (\frac{\chi}{10^{-2}})^{0.5}$, where $\chi$ is the suppression factor of the diffusion coefficient near the source compared with the Galactic diffusion coefficient, $D_{source}(E)=\chi D(E)$. The angular size of the \gray emission is estimated as $r/d \sim 0.33^{\circ}(\frac{\chi}{10^{-2}})^{0.5}$.  The maximum diffusion length, in any case,  should be smaller than 110~pc,  the ballistic propagation length  over 340 years.  And the maximum angular extension of the \gray emission when CRs are in the diffusive regime should be smaller than $1.8^{\circ}$. To estimate the total CR energy budget from the \gray luminosity, only the average density matters. As estimated in \citet{ma19}, the average density in the nearby 100 pc of Cas A  is  $10~\rm cm^{-3}$. In this case, due to the lower average density and larger integration area in deriving the \gray flux, the derived CR energy budget would be more conservative than the ballistic propagation case we considered above. To derive the robust and model independent estimations we consider only the diffusive regime in the following discussion.

%According to the different sizes of the possible  $\gamma$-ray emission region, we estimated the total mass  separately based on the gas column density map as shown in Figure.\ref{fig:co}. For ballistic diffusion case, we used the size of the SNR, which is about $0.04^{\circ}$ in radius, the total mass is $1100~\rm M_{\odot}$.  For $\chi \sim 10^{-4}$ the prpgation length is of  about $2 ~\rm pc$, which is similar to the size of the SNRs and we use the same $1100~\rm M_{\odot}$ in this case. Assuming a spherical symmetric geometry the average density is about $1000 ~\rm cm^{-3}$. Note that this should be regarded as an upper limit, since we indeed integrate all the gas in the velocity interval in the line of sight, considering the size of the GMC is larger than $2 ~\rm pc$, it is quite likely a significant of the gas are in the foreground/background.   For $\chi \sim 10^{-2}$, the propagation length is about $20~\rm pc$,  the total mass is about $5\times 10^4 ~\rm M_{\odot}$ and the average density is $50 ~\rm cm^{-3}$. And for the most conservative case, all the molecular gas has been illuminated by the escaping CRs, the total mass is estimated above as $9.5\times 10^5~\rm M_{\odot}$ with average density of $10~\rm cm^{-3}$.

%and the average density is estimated as $170 \rm cm^{-3}$

%In this case, the VHE proton occupied region should be a point source in the view of LHAASO KM2A. 

\section{KM2A data analysis} \label{sec:data}
The data used in this analysis were collected from December 2019 to September 2022 by the half KM2A, three quarters KM2A and full KM2A. Quality selections have been applied to ensure a reliable data quality.  We select the periods when at least 95$\%$ of working sub-detectors were in good condition. The total effective observation time is 932.74 days after data quality selection. To achieve a better performance of the array, several event cuts are also implemented. The cuts on data and event selections are the same as listed in the KM2A performance paper \citep{2021ChPhC..45b5002A}.

Considering the energy resolution and statistics, one decade of energy is divided into 5 bins with a bin width of $log_{10}E = 0.2$. The sky in celestial coordinates (right ascension and declination) is divided into cells of size $0.1^{\circ}\times0.1^{\circ}$ and filled with detected events according to their reconstructed arrival directions for each energy bin. To extract the excess of $\gamma$-rays from each cell, the ``direct integration method'' \citep{2004ApJ...603..355F} is adopted to estimate the number of cosmic ray background events. The test statistic (TS) used to evaluate the significance of the source in this work is ${\rm TS} = 2\log(\lambda)$, where $\lambda={\mathcal{L}_{1}/\mathcal{L}_0}$, $\mathcal{L}_{1}$ is the maximum likelihood value for the alternative hypothesis, and ${\mathcal{L}_0}$ is the maximum likelihood value of the null hypothesis. The TS map of the ROI is shown in the left panel of Fig.\ref{fig:co}. We found that there is no significant excess in the vicinity of Cas A. 
%The uniform disk model is used in this analysis.  The disk model is convolved with the angular resolution of detector. Two different radius , $0.3^{\circ}$ and $2.0^{\circ}$, are assumed. 
%{\bf As mentioned in the last section, we consdier both the ballistic propagation and diffusion regime. In the former case, the expected \gray emission produced by CRs escaped from Cas A are point-like. In KM2A analysis, taken into account the point spread function (psf), we integrate all the photons in the a circle of $0.6^{\circ}$ (in which 90\% of the photons from a point source are included). For diffusion regime, the intrinsic size of the source varies as the diffusion coefficient, however, the maximum angular radius is $1.8^{\circ}$ as estimated in last section. Since the upper limit of \gray flux grow as the integration size, any upper limit from smaller size would give more stringent constraints.  Taken into account of the psf, we use an integration radius  of $2.4^{\circ}$ to estimate the \gray upper limit, which is robust and conservative.  }

\begin{figure}
    \centering
    \includegraphics{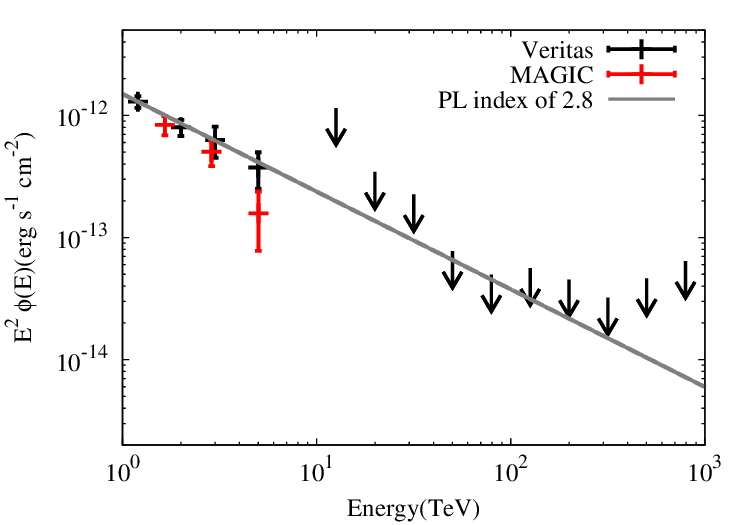}
    \caption{The UHE $\gamma$-ray flux upper limits assuming Cas A is a point source. The data points are from MAGIC and VERITAS  obervations.  The grey curve is the power law function with the index of 2.8. A harder spectrum would violate the LHAASO upper limit.  }
    \label{fig:upper_ps}
\end{figure}

\begin{figure}
    \centering
    \includegraphics{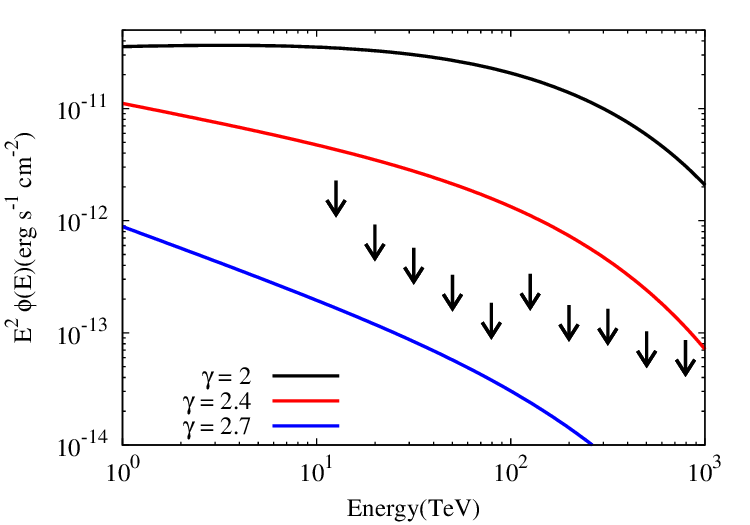}
    \caption{The UHE $\gamma$-ray flux upper limits within 1.8 degree from Cas A.  The black, red and blue curves correspond to the predicted pion-decay $\gamma$-ray fluxes for the proton index of 2.0, 2.4 and 2.7, respectively. The total energies of CR protons within $100 - 1000~\rm TeV$ are $5\times 10^{49} \rm erg$, $3\times 10^{48} \rm erg$ and $1\times 10^{47} \rm erg$, respectively, which are the CR energy budgets expected if Cas A type SNR produce all the PeV CRs in our Galaxy. }
    \label{fig:upper_ext}
\end{figure}

As mentioned in the last section, we consider both the ballistic propagation and diffusive regime. In the former case, the expected \gray emission produced by CRs escaping from Cas~A is point-like. In KM2A analysis, taking into account the point spread function (PSF), we integrate all photons in the 90\% contamination radius  from a point source. For diffusive regime, the intrinsic size of the source varies with the diffusion coefficient; however, the maximum angular radius is 1.8$^\circ$ as estimated in the previous section. Taking into account of the angular resolution of the detector, we use an integration radius of 2.2$^\circ$ for the lower energy bin and 1.85$^\circ$ for energies higher than 100 TeV, to estimate the upper limits. No significant excess has been  detected. We derived an upper limit for each energy bin at the 95\% confidence level using the method proposed by \citet{HELENE1983319}. For a more conservative estimate of the upper limit, we use the CERN ROOT tools following \citet{Feldman-Cousins1998} to calculate the 95\% upper limit of the signal when the background count is less than 20. The results are shown in Fig.\ref{fig:upper_ps} and Fig.\ref{fig:upper_ext}. The main uncertainties come from the  uncertainty in the differential spectral index of the \gray emission,  which we  assume to derive  the upper limit. In order to estimate such systematic uncertainties, the analysis is separately performed for the \gray spectral index of -2 or -3. The uncertainties found from this test are small: about 10\% and 3\% for the first two energy bins, and less than 1\% at higher energies.

 The systematic uncertainties affecting the flux upper limits are similar to those studied in \citet{2021ChPhC..45b5002A}. The number of operating units varied with time due to debugging a few percent of detector units. The variable layout of the array, affecting the $\gamma$-ray/background separation, has a slight effect on the flux. The systematic uncertainty also comes from the atmospheric model used in the simulation, which affects the detection efficiency. The overall systematic uncertainty affecting the flux is estimated to be $\sim$7$\%$. 

In principle, the diffuse galactic \gray emission (DGE) should be subtracted before we derive the upper limit for sources. But the inclusion of such a component in deriving the upper limit only has a marginal (10\%) impact on the final results. Thus to avoid further systematic errors, we do not subtract the DGE in this work and use the most conservative upper limits.

%% The "ht!" tells LaTeX to put the figure "here" first, at the "top" next
%% and to override the normal way of calculating a float position

\section{Discussion}
\subsection{Point source scenario}
The recent MAGIC and VERITAS observations reveal a significant steepening
 in the $\gamma$-ray spectrum at TeV energies \citep{casa_magic, Abeysekara_2020}.  This can be interpreted as a break or a cutoff in the spectrum. The latter would imply that Cas A  accelerates  particles only up to tens of TeV.  However,  recent studies (see e.g.  \cite{Malkov_steep}) show that taking into account realistic  geometry and magnetic field configurations,  one can expect  a significant deviation from the standard DSA predictions, namely very steep,  power-law type  spectra. This option  does not need  a cutoff to describe  the TeV data; a broken power law can fit the   GeV-TeV SEDs well, with a spectral index of about 2.7 above the break \citep{Abeysekara_2020}. 
 
 To explore whether this scenario is compatible with the KM2A observations, we extrapolate the data points measured by IACTs with a power law. It should be noted that in this scenario Cas A can still be in the PeVatron phase, and thus the gamma-ray observations cannot set a limit on the total CR energy budget above $100~\rm TeV$. But information regarding the acceleration spectrum can be obtained from the observed \gray spectrum.  We found that a spectral index smaller than 2.8 already violates the KM2A upper limit. The \gray spectral index of 2.8 corresponds to the index of  $\simeq 2.9$ for CR protons. Such an injection spectrum is softer than the CR spectrum measured in this energy range,  even ignoring the   further steepening  of the  spectrum of  particles during their propagation in the galactic plane and halo.  Such a soft injection spectrum can hardly be accommodated in the current  CR propagation  concept; thus,  their contribution to the locally measured CR flux can not be significant.

\subsection{ Extended source scenario }
 One should note that, in general, independent of the question of its contribution to the local CR flux, the very steep proton spectrum  does not exclude acceleration of PeV protons by Cas A.  Therefore, to probe the amount of   PeV protons accelerated at the early epochs  of  Cas~A,  we assume an integrated spectrum  of these CRs in the form  $N(E) \sim E^{-\gamma} {e^{-E/E_c}}$. Since we are considering the possibility of Cas A as a PeVatron, we fix $E_c$ =2$~\rm PeV$,  and  choose the index $\gamma=$2.0, 2.4, or 2.7 (see the discussion below) .  As long as we are investigating CRs released between $100~\rm TeV$ and  $1~\rm PeV$, the exact value of the cutoff energy does not have a strong  impact on the total energy budget. %We also ignore the KM2A upper limit above $100~\rm TeV$ in the point source case; the \gray spectrum above $100~\rm TeV$  strongly depends on the exact cutoff shape of the CR spectrum, and the inclusion of these points would induce more stringent constraints on the total CR energy budget.
The $\gamma$-ray flux is proportional to the product of CR total energy and gas density. Thus from the observed upper limit of $\gamma$-ray flux and the derived gas density, one can derive an upper limit for the CR total energy. By using the $\gamma$-ray production cross-section parametrised in \citet{Kafexhiu14},  we calculated the predicted $\gamma$-ray flux for $\gamma=$ 2.0, 2.4, or 2.7. Taking into account the average gas density of about $10~\rm cm^{-3}$, we derived the upper limit of total energy of CRs in the energy range 100 - 1000 $~\rm TeV$ as 3.6/3.0/4.0$\times 10^{47}$ (10 cm$^{-3}$/$n$) $~\rm erg$, assuming all the molecular gas has been illuminated by the escaped CRs, with an index of 2.0/2.4/2.7.  The derived upper limit of total CR energy only slightly depends on the spectral index. 

If the appearance of  the Cas~A-like SNRs events has a rate of 1 per century, as suggested by \citet{schure13}, the  CR injection rate in our Galaxy  in the energy interval $100-1000~\rm TeV$ would be   $1.2/1.0/1.3 \times 10^{38} ~\rm erg/s$ for an injection index of 2.0/2.4/2.7, respectively. The total CR injection rate is estimated as $1 - 3 \times 10^{41} ~\rm erg/s$ in our Galaxy \citep{Drury12}, which corresponds to an injection rate in the energy interval $100-1000~\rm TeV$ as $1.6 - 5.0 \times 10^{40} ~\rm erg/s$ for $\gamma=2.0$, $1.0 - 3.0 \times 10^{39} ~\rm erg/s$ for $\gamma=2.4$ and $3.0 - 9.0 \times 10^{37} ~\rm erg/s$ for $\gamma=2.7$. %As an alternative way to express this result, we plotted the predicted \gray flux in Figure \ref{fig:upper_ext} assuming the CRs injected by Cas A type SNRs dominated the PeV CRs in our Galaxy. %We also plotted the predicted \gray flux in Figure .\ref{fig:upper_ext} assuming the CRs injected by Cas A type SNRs dominated the PeV CRs in our Galaxy. 
For the acceleration spectral index of 2.0, which is predicted in the standard DSA theory, the Cas A upper limits are more than two orders of magnitude smaller than that required for the total CR energy budget in our Galaxy. For a softer injection spectral index of 2.4,  the $100-1000~\rm TeV$ CR injection power from Cas A type SNRs  derived here is  still one order of magnitude smaller than the requirement for the whole Galaxy. And in the case of the  injection spectral  index of 2.7, we cannot rule out the possibility that all CRs above $100~\rm TeV$ in our Galaxy are injected from Cas A type objects.  As an alternative way to express this result, we plotted the predicted \gray flux in Figure \ref{fig:upper_ext} assuming the CRs injected by Cas A type SNRs dominated the PeV CRs in our Galaxy. We again see that the limits from LHAASO KM2A rule out that Cas A type SNRs produce all PeV CRs in the Galaxy for $\gamma = $ 2.0 or 2.4,but do not rule out if $\gamma = $ 2.7.

Indeed, as shown in the last paragraph, the upper limit of the rate of  CR  above $100~\rm TeV$ injected by Cas A only slightly depends on the injection spectrum and is estimated as $\sim 1.3 \times 10^{38} (10~\rm cm^{-3}/n) ~\rm erg/s$. On the other hand, the total energy budget required for CRs above $100~\rm TeV$ depends only on the injection spectrum. If we assume a pure power law, the injection energy budget above $100~\rm TeV$ is $W(E>100~\rm TeV)=  (1 - 3) \times 10^{(41-5(\Gamma-2))} ~\rm erg/s$ for $\Gamma>2.0$, where $\Gamma$ is the injection spectral index.  By comparing with the limit set by KM2A observations on Cas A, we found that for $n \simeq 10~\rm cm^{-3}$  as estimated from CO observations, our results can rule out the hypothesis that Cas A type SNRs are  the only PeVatrons contributing to CRs above  $100~\rm TeV$ for a CR injection spectrum index less than $2.5$. A smaller gas density would relax the constraints. But note that an average gas density  well  below $10 ~\rm cm^3$ seems unrealistic in this region. In the most conservative case, assuming a gas density of $1 ~\rm cm^{-3}$, the lower limit for the  injection spectral index is $2.4$. 

  In scenario (3), the highest energy CRs have already escaped from the shock upstream. Cas A still lies in  its progenitor's wind bubble \citep{weil20},  which has a size of about $10~\rm pc$ and a much lower density compared with the ISM. In the most extreme case, it is possible that CRs escaped from the shell and confined in the bubble.  The angular size of the bubble is about 5 arcminutes, which is much smaller than the PSF of KM2A. As a result, the bubble is also a point source in KM2A, and the upper limit derived from a point source is about 1/5 of the extended source scenario considered above.  If we assume an average density of $0.1~\rm cm^{-3}$ in the wind bubble, the constraints on the total CR energy budget can be 20 times larger than our estimation above.

In the derivation of the \gray upper limit, the Galactic diffuse \gray emission is not considered, which would make the upper limit even lower and the constraint more stringent. Thus we argue that our calculation implies that the Cas A-type SNRs cannot be the only sources for CRs up to PeV in the regime of standard DSA (hard injection spectrum). If we assume a softer injection spectrum, Cas A type SNRs can still be major PeVatrons in our Galaxy. But such a softer injection spectrum would be  in tension with most CR propagation models, in which a strong energy dependence of the CR diffusion coefficient is used, and the CR injection spectrum should be hard to fit the locally observed CR spectrum. 

We also note that a more complex spectral shape, such as a broken power law,  is possible for the total CR flux injected by Cas A. But in this work our constraints mainly come from a rather narrow energy band, and the constraints on the exact spectral shape are limited. To avoid uncertainties and degeneracy, we used a single power law spectra in this study.

In conclusion,  based on the LHAASO KM2A measurements, we set stringent upper limits on the total energy budget of  UHE protons accelerated by Cas A. Our constraints here are nearly model independent, unless in the hypothetical, and in our view, not very realistic situation, when cosmic rays are confined in a compact low density wind bubble after escaping from the shell. 
Although we cannot formally rule out the possibility that the Cas A type SNRs are the Galactic PeVatrons,  the stringent limits on the accelerated (injection) spectrum have important implications. 
%For soft accelerated spectra (softer than 2.4 or 2.5) we can already rule out specific models such as the magnetic amplification by NRH instability \citep{schure13}. 
Further UHE observations will investigate more parameter spaces. We note that the most stringent limits come from the \gray fluxes above $100~\rm TeV$, for which the sensitivity of LHAASO KM2A will remain unrivalled in the foreseeable  future. We anticipate  that the accumulation of LHAASO KM2A data for another 5 to 10 years will improve the upper limit substantially  and would finally rule out or confirm Cas A  type SNRs as a PeVatron population. 

%Thus another population of PeV CRs are required to account for the observed CR spectrum. Or the CR halo size is much smaller or the confinement time of PeV protons are much longer than expected. 

%\section{Conclusions}

\acknowledgments

%%%We would like to thank all staff members who work at the LHAASO site above 4400 meters above sea level year-round to maintain the detector and keep the water recycling system, electricity power supply and other components of the experiment operating smoothly. We are grateful to Chengdu Management Committee of Tianfu New Area for the constant financial support for research with LHAASO data. This research work is also supported by the following grants: The National Key R&D program of China under grants 2018YFA0404201, 2018YFA0404202, 2018YFA0404203, and 2018YFA0404204, by the National Natural Science Foundation of China  NSFC No.12022502, No.12205314, No. 12105301, No. 12261160362, No.12105294, No.U1931201. In Thailand, support was provided by the National Science and Technology Development Agency (NSTDA) and the National Research Council of Thailand (NRCT) under the High-Potential Research Team Grant Program (N42A650868). 
We would like to thank all staff members who work at the LHAASO site above 4400 meters above sea level year-round to maintain the detector and keep the water recycling system, electricity power supply and other components of the experiment operating smoothly. We are grateful to Chengdu Management Committee of Tianfu New Area for the constant financial support for research with LHAASO data. We deeply appreciate the computing and data service support provided by the National High Energy Physics Data Center for the data analysis in this paper. This research work is also supported by the following grants: The National Key R\&D program of China under grants 2018YFA0404201, 2018YFA0404202, 2018YFA0404203, and 2018YFA0404204, by the National Natural Science Foundation of China NSFC No.12022502, No.12205314, No. 12105301, No. 12261160362, No.12105294, No.U1931201. In Thailand, support was provided by the National Science and Technology Development Agency (NSTDA) and the National Research Council of Thailand (NRCT) under the High-Potential Research Team Grant Program (N42A650868).

\bibliography{casa}{}

\begin{thebibliography}{}
\expandafter\ifx\csname natexlab\endcsname\relax\def\natexlab#1{#1}\fi
\providecommand{\url}[1]{\href{#1}{#1}}
\providecommand{\dodoi}[1]{doi:~\href{http://doi.org/#1}{\nolinkurl{#1}}}
\providecommand{\doeprint}[1]{\href{http://ascl.net/#1}{\nolinkurl{http://ascl.net/#1}}}
\providecommand{\doarXiv}[1]{\href{https://arxiv.org/abs/#1}{\nolinkurl{https://arxiv.org/abs/#1}}}

\bibitem[{Abeysekara {et~al.}(2020)Abeysekara, Archer, Benbow, Bird, Brose,
  Buchovecky, Buckley, Chromey, Cui, Daniel, Das, Dwarkadas, Falcone, Feng,
  Finley, Fortson, Gent, Gillanders, Giuri, Gueta, Hanna, Hassan, Hervet,
  Holder, Hughes, Humensky, Kaaret, Kar, Kelley-Hoskins, Kertzman, Kieda,
  Krause, Krennrich, Kumar, Lang, Maier, Moriarty, Mukherjee, Nievas-Rosillo,
  O’Brien, Ong, Park, Petrashyk, Pfrang, Pohl, Pueschel, Quinn, Ragan,
  Reynolds, Richards, Roache, Sadeh, Santander, Sembroski, Shahinyan, Sushch,
  Weinstein, Wilcox, Wilhelm, Williams, Williamson, Zitzer, \&
  Ghiotto}]{Abeysekara_2020}
Abeysekara, A.~U., Archer, A., Benbow, W., {et~al.} 2020, The Astrophysical
  Journal, 894, 51, \dodoi{10.3847/1538-4357/ab8310}

\bibitem[{{Acciari} {et~al.}(2010){Acciari}, {Aliu}, {Arlen}, {Aune},
  {Bautista}, {Beilicke}, {Benbow}, {Boltuch}, {Bradbury}, {Buckley}, {Bugaev},
  {Butt}, {Byrum}, {Cannon}, {Cesarini}, {Chow}, {Ciupik}, {Cogan}, {Cui},
  {Dickherber}, {Duke}, {Ergin}, {Fegan}, {Finley}, {Finnegan}, {Fortin},
  {Fortson}, {Furniss}, {Galante}, {Gall}, {Gillanders}, {Grube}, {Guenette},
  {Gyuk}, {Hanna}, {Holder}, {Huang}, {Hui}, {Humensky}, {Kaaret}, {Karlsson},
  {Kertzman}, {Kieda}, {Konopelko}, {Krawczynski}, {Krennrich}, {Lang},
  {LeBohec}, {Maier}, {McArthur}, {McCann}, {McCutcheon}, {Millis}, {Moriarty},
  {Ong}, {Pandel}, {Perkins}, {Pohl}, {Quinn}, {Ragan}, {Reynolds}, {Roache},
  {Rose}, {Schroedter}, {Sembroski}, {Smith}, {Smith}, {Steele}, {Swordy},
  {Theiling}, {Thibadeau}, {Varlotta}, {Vassiliev}, {Vincent}, {Wagner},
  {Wakely}, {Ward}, {Weekes}, {Weinstein}, {Weisgarber}, {Wissel}, {Wood}, \&
  {VERITAS Collaboration}}]{casa_veritas}
{Acciari}, V.~A., {Aliu}, E., {Arlen}, T., {et~al.} 2010, \apj, 714, 163,
  \dodoi{10.1088/0004-637X/714/1/163}

\bibitem[{{Acciari} {et~al.}(2011){Acciari}, {Aliu}, {Arlen}, {Aune},
  {Beilicke}, {Benbow}, {Bradbury}, {Buckley}, {Bugaev}, {Byrum}, {Cannon},
  {Cesarini}, {Ciupik}, {Collins-Hughes}, {Cui}, {Dickherber}, {Duke},
  {Errando}, {Finley}, {Finnegan}, {Fortson}, {Furniss}, {Galante}, {Gall},
  {Gillanders}, {Godambe}, {Griffin}, {Grube}, {Guenette}, {Gyuk}, {Hanna},
  {Holder}, {Hughes}, {Hui}, {Humensky}, {Kaaret}, {Karlsson}, {Kertzman},
  {Kieda}, {Krawczynski}, {Krennrich}, {Lang}, {LeBohec}, {Madhavan}, {Maier},
  {Majumdar}, {McArthur}, {McCann}, {Moriarty}, {Mukherjee}, {Ong}, {Orr},
  {Otte}, {Pandel}, {Park}, {Perkins}, {Pohl}, {Quinn}, {Ragan}, {Reyes},
  {Reynolds}, {Roache}, {Rose}, {Saxon}, {Schroedter}, {Sembroski}, {Senturk},
  {Slane}, {Smith}, {Te{\v{s}}i{\'c}}, {Theiling}, {Thibadeau}, {Tsurusaki},
  {Varlotta}, {Vassiliev}, {Vincent}, {Vivier}, {Wakely}, {Ward}, {Weekes},
  {Weinstein}, {Weisgarber}, {Williams}, {Wood}, \& {Zitzer}}]{tycho_veritas}
---. 2011, \apjl, 730, L20, \dodoi{10.1088/2041-8205/730/2/L20}

\bibitem[{{Aharonian} {et~al.}(2019){Aharonian}, {Yang}, \& {de O{\~n}a
  Wilhelmi}}]{aharonian19}
{Aharonian}, F., {Yang}, R., \& {de O{\~n}a Wilhelmi}, E. 2019, Nature
  Astronomy, 3, 561, \dodoi{10.1038/s41550-019-0724-0}

\bibitem[{{Aharonian} {et~al.}(2001){Aharonian}, {Akhperjanian}, {Barrio},
  {Bernl{\"o}hr}, {B{\"o}rst}, {Bojahr}, {Bolz}, {Contreras}, {Cortina},
  {Denninghoff}, {Fonseca}, {Gonzalez}, {G{\"o}tting}, {Heinzelmann},
  {Hermann}, {Heusler}, {Hofmann}, {Horns}, {Ibarra}, {Iserlohe}, {Jung},
  {Kankanyan}, {Kestel}, {Kettler}, {Kohnle}, {Konopelko}, {Kornmeyer},
  {Kranich}, {Krawczynski}, {Lampeitl}, {Lopez}, {Lorenz}, {Lucarelli},
  {Magnussen}, {Mang}, {Meyer}, {Mirzoyan}, {Moralejo}, {Ona}, {Padilla},
  {Panter}, {Plaga}, {Plyasheshnikov}, {Prahl}, {P{\"u}hlhofer}, {Rauterberg},
  {R{\"o}hring}, {Rhode}, {Rowell}, {Sahakian}, {Samorski}, {Schilling},
  {Schr{\"o}der}, {Siems}, {Stamm}, {Tluczykont}, {V{\"o}lk}, {Wiedner}, \&
  {Wittek}}]{casa_hegra}
{Aharonian}, F., {Akhperjanian}, A., {Barrio}, J., {et~al.} 2001, \aap, 370,
  112, \dodoi{10.1051/0004-6361:20010243}

\bibitem[{{Aharonian} {et~al.}(2021){Aharonian}, {An}, {Axikegu}, {Bai}, {Bai},
  {Bao}, {Bastieri}, {Bi}, {Bi}, {Cai}, {Cai}, {Cao}, {Cao}, {Chang}, {Chang},
  {Chang}, {Chen}, {Chen}, {Chen}, {Chen}, {Chen}, {Chen}, {Chen}, {Chen},
  {Chen}, {Chen}, {Chen}, {Chen}, {Chen}, {Cheng}, {Cheng}, {Cui}, {Cui},
  {Cui}, {Dai}, {Dai}, {Dai}, {Danzengluobu}, {Della Volpe}, {Piazzoli},
  {Dong}, {Fan}, {Fan}, {Fan}, {Fang}, {Fang}, {Feng}, {Feng}, {Feng}, {Feng},
  {Gao}, {Gao}, {Gao}, {Gao}, {Ge}, {Geng}, {Gong}, {Gou}, {Gu}, {Guo}, {Guo},
  {Guo}, {Guo}, {Han}, {He}, {He}, {He}, {He}, {He}, {He}, {Heller}, {Hor},
  {Hou}, {Hou}, {Hu}, {Hu}, {Hu}, {Hu}, {Huang}, {Huang}, {Huang}, {Huang},
  {Huang}, {Ji}, {Ji}, {Jia}, {Jiang}, {Jiang}, {Jin}, {Kuleshov}, {Levochkin},
  {Li}, {Li}, {Li}, {Li}, {Li}, {Li}, {Li}, {Li}, {Li}, {Li}, {Li}, {Li}, {Li},
  {Li}, {Li}, {Li}, {Li}, {Liang}, {Liang}, {Lin}, {Liu}, {Liu}, {Liu}, {Liu},
  {Liu}, {Liu}, {Liu}, {Liu}, {Liu}, {Liu}, {Liu}, {Liu}, {Liu}, {Liu}, {Liu},
  {Long}, {Lu}, {Lv}, {Ma}, {Ma}, {Ma}, {Mao}, {Masood}, {Mitthumsiri},
  {Montaruli}, {Nan}, {Pang}, {Pattarakijwanich}, {Pei}, {Qi}, {Ruffolo},
  {Rulev}, {S{\'a}iz}, {Shao}, {Shchegolev}, {Sheng}, {Shi}, {Song}, {Stenkin},
  {Stepanov}, {Sun}, {Sun}, {Sun}, {Tam}, {Tang}, {Tian}, {Wang}, {Wang},
  {Wang}, {Wang}, {Wang}, {Wang}, {Wang}, {Wang}, {Wang}, {Wang}, {Wang},
  {Wang}, {Wang}, {Wang}, {Wang}, {Wang}, {Wang}, {Wang}, {Wang}, {Wang},
  {Wang}, {Wei}, {Wei}, {Wei}, {Wen}, {Wu}, {Wu}, {Wu}, {Wu}, {Wu}, {Xi},
  {Xia}, {Xia}, {Xiang}, {Xiao}, {Xiao}, {Xin}, {Xin}, {Xing}, {Xu}, {Xu},
  {Xue}, {Yan}, {Yang}, {Yang}, {Yang}, {Yang}, {Yang}, {Yang}, {Yang}, {Yao},
  {Yao}, {Ye}, {Yin}, {Yin}, {You}, {You}, {Yu}, {Yuan}, {Zeng}, {Zeng},
  {Zeng}, {Zeng}, {Zha}, {Zhai}, {Zhang}, {Zhang}, {Zhang}, {Zhang}, {Zhang},
  {Zhang}, {Zhang}, {Zhang}, {Zhang}, {Zhang}, {Zhang}, {Zhang}, {Zhang},
  {Zhang}, {Zhang}, {Zhang}, {Zhang}, {Zhang}, {Zhang}, {Zhao}, {Zhao}, {Zhao},
  {Zhao}, {Zhao}, {Zheng}, {Zheng}, {Zhou}, {Zhou}, {Zhou}, {Zhou}, {Zhou},
  {Zhou}, {Zhu}, {Zhu}, {Zhu}, {Zhu}, {Zuo}, \& {(Lhaaso
  Collaboration)}}]{2021ChPhC..45b5002A}
{Aharonian}, F., {An}, Q., {Axikegu}, {et~al.} 2021, Chinese Physics C, 45,
  025002, \dodoi{10.1088/1674-1137/abd01b}

\bibitem[{{Ahnen} {et~al.}(2017){Ahnen}, {Ansoldi}, {Antonelli}, {Arcaro},
  {Babi{\'c}}, {Banerjee}, {Bangale}, {Barres de Almeida}, {Barrio}, {Becerra
  Gonz{\'a}lez}, {Bednarek}, {Bernardini}, {Berti}, {Bhattacharyya},
  {Biasuzzi}, {Biland}, {Blanch}, {Bonnefoy}, {Bonnoli}, {Carosi}, {Carosi},
  {Chatterjee}, {Colak}, {Colin}, {Colombo}, {Contreras}, {Cortina}, {Covino},
  {Cumani}, {Da Vela}, {Dazzi}, {De Angelis}, {De Lotto}, {de O{\~n}a
  Wilhelmi}, {Di Pierro}, {Doert}, {Dom{\'\i}nguez}, {Dominis Prester},
  {Dorner}, {Doro}, {Einecke}, {Eisenacher Glawion}, {Elsaesser},
  {Engelkemeier}, {Fallah Ramazani}, {Fern{\'a}ndez-Barral}, {Fidalgo},
  {Fonseca}, {Font}, {Fruck}, {Galindo}, {Garc{\'\i}a L{\'o}pez},
  {Garczarczyk}, {Gaug}, {Giammaria}, {Godinovi{\'c}}, {Gora}, {Guberman},
  {Hadasch}, {Hahn}, {Hassan}, {Hayashida}, {Herrera}, {Hose}, {Hrupec},
  {Inada}, {Ishio}, {Konno}, {Kubo}, {Kushida}, {Kuve{\v{z}}di{\'c}}, {Lelas},
  {Lindfors}, {Lombardi}, {Longo}, {L{\'o}pez}, {Maggio}, {Majumdar},
  {Makariev}, {Maneva}, {Manganaro}, {Mannheim}, {Maraschi}, {Mariotti},
  {Mart{\'\i}nez}, {Mazin}, {Menzel}, {Minev}, {Mirzoyan}, {Moralejo},
  {Moreno}, {Moretti}, {Neustroev}, {Niedzwiecki}, {Nievas Rosillo}, {Nilsson},
  {Ninci}, {Nishijima}, {Noda}, {Nogu{\'e}s}, {Paiano}, {Palacio}, {Paneque},
  {Paoletti}, {Paredes}, {Pedaletti}, {Peresano}, {Perri}, {Persic}, {Prada
  Moroni}, {Prandini}, {Puljak}, {Garcia}, {Reichardt}, {Rhode}, {Rib{\'o}},
  {Rico}, {Righi}, {Saito}, {Satalecka}, {Schroeder}, {Schweizer}, {Shore},
  {Sitarek}, {{\v{S}}nidari{\'c}}, {Sobczynska}, {Stamerra}, {Strzys},
  {Suri{\'c}}, {Takalo}, {Tavecchio}, {Temnikov}, {Terzi{\'c}}, {Tescaro},
  {Teshima}, {Torres-Alb{\`a}}, {Treves}, {Vanzo}, {Vazquez Acosta}, {Vovk},
  {Ward}, {Will}, \& {Zari{\'c}}}]{casa_magic}
{Ahnen}, M.~L., {Ansoldi}, S., {Antonelli}, L.~A., {et~al.} 2017, \mnras, 472,
  2956, \dodoi{10.1093/mnras/stx2079}

\bibitem[{{Bell}(2004)}]{Bell2004}
{Bell}, A.~R. 2004, \mnras, 353, 550, \dodoi{10.1111/j.1365-2966.2004.08097.x}

\bibitem[{{Bell} {et~al.}(2019){Bell}, {Matthews}, \& {Blundell}}]{Bell_steep}
{Bell}, A.~R., {Matthews}, J.~H., \& {Blundell}, K.~M. 2019, \mnras, 488, 2466,
  \dodoi{10.1093/mnras/stz1805}

\bibitem[{{Bell} {et~al.}(2011){Bell}, {Schure}, \& {Reville}}]{bell11}
{Bell}, A.~R., {Schure}, K.~M., \& {Reville}, B. 2011, \mnras, 418, 1208,
  \dodoi{10.1111/j.1365-2966.2011.19571.x}

\bibitem[{{Bell} {et~al.}(2013){Bell}, {Schure}, {Reville}, \&
  {Giacinti}}]{bell13}
{Bell}, A.~R., {Schure}, K.~M., {Reville}, B., \& {Giacinti}, G. 2013, \mnras,
  431, 415, \dodoi{10.1093/mnras/stt179}

\bibitem[{{Bolatto} {et~al.}(2013){Bolatto}, {Wolfire}, \& {Leroy}}]{bolatto13}
{Bolatto}, A.~D., {Wolfire}, M., \& {Leroy}, A.~K. 2013, \araa, 51, 207,
  \dodoi{10.1146/annurev-astro-082812-140944}

\bibitem[{{Caprioli} {et~al.}(2020){Caprioli}, {Haggerty}, \&
  {Blasi}}]{Caprioli_steep}
{Caprioli}, D., {Haggerty}, C.~C., \& {Blasi}, P. 2020, \apj, 905, 2,
  \dodoi{10.3847/1538-4357/abbe05}

\bibitem[{{Drury}(2012)}]{Drury12}
{Drury}, L.~O.~. 2012, Astroparticle Physics, 39, 52,
  \dodoi{10.1016/j.astropartphys.2012.02.006}

\bibitem[{Feldman \& Cousins(1998)}]{Feldman-Cousins1998}
Feldman, G.~J., \& Cousins, R.~D. 1998, Phys. Rev. D, 57, 3873,
  \dodoi{10.1103/PhysRevD.57.3873}

\bibitem[{{Fleysher} {et~al.}(2004){Fleysher}, {Fleysher}, {Nemethy}, {Mincer},
  \& {Haines}}]{2004ApJ...603..355F}
{Fleysher}, R., {Fleysher}, L., {Nemethy}, P., {Mincer}, A.~I., \& {Haines},
  T.~J. 2004, \apj, 603, 355, \dodoi{10.1086/381384}

\bibitem[{{Grefenstette} {et~al.}(2015){Grefenstette}, {Reynolds}, {Harrison},
  {Humensky}, {Boggs}, {Fryer}, {DeLaney}, {Madsen}, {Miyasaka}, {Wik},
  {Zoglauer}, {Forster}, {Kitaguchi}, {Lopez}, {Nynka}, {Christensen}, {Craig},
  {Hailey}, {Stern}, \& {Zhang}}]{casa_nustar}
{Grefenstette}, B.~W., {Reynolds}, S.~P., {Harrison}, F.~A., {et~al.} 2015,
  \apj, 802, 15, \dodoi{10.1088/0004-637X/802/1/15}

\bibitem[{{Hanusch} {et~al.}(2019){Hanusch}, {Liseykina}, {Malkov}, \&
  {Aharonian}}]{Malkov_steep2}
{Hanusch}, A., {Liseykina}, T.~V., {Malkov}, M., \& {Aharonian}, F. 2019, \apj,
  885, 11, \dodoi{10.3847/1538-4357/ab426d}

\bibitem[{Helene(1983)}]{HELENE1983319}
Helene, O. 1983, Nuclear Instruments and Methods in Physics Research, 212, 319,
  \dodoi{https://doi.org/10.1016/0167-5087(83)90709-3}

\bibitem[{{Kafexhiu} {et~al.}(2014){Kafexhiu}, {Aharonian}, {Taylor}, \&
  {Vila}}]{Kafexhiu14}
{Kafexhiu}, E., {Aharonian}, F., {Taylor}, A.~M., \& {Vila}, G.~S. 2014, \prd,
  90, 123014, \dodoi{10.1103/PhysRevD.90.123014}

\bibitem[{{Kassim} {et~al.}(1995){Kassim}, {Perley}, {Dwarakanath}, \&
  {Erickson}}]{kassim95}
{Kassim}, N.~E., {Perley}, R.~A., {Dwarakanath}, K.~S., \& {Erickson}, W.~C.
  1995, \apjl, 455, L59, \dodoi{10.1086/309802}

\bibitem[{{Kim} {et~al.}(2008){Kim}, {Rieke}, {Krause}, {Misselt},
  {Indebetouw}, \& {Johnson}}]{casa_ism}
{Kim}, Y., {Rieke}, G.~H., {Krause}, O., {et~al.} 2008, \apj, 678, 287,
  \dodoi{10.1086/533426}

\bibitem[{{Krause} {et~al.}(2008){Krause}, {Birkmann}, {Usuda}, {Hattori},
  {Goto}, {Rieke}, \& {Misselt}}]{kraus08}
{Krause}, O., {Birkmann}, S.~M., {Usuda}, T., {et~al.} 2008, Science, 320,
  1195, \dodoi{10.1126/science.1155788}

\bibitem[{{Lagage} \& {Cesarsky}(1983)}]{LagageCesarsky}
{Lagage}, P.~O., \& {Cesarsky}, C.~J. 1983, \aap, 125, 249

\bibitem[{{Lazarian} \& {Xu}(2021)}]{lazarian21}
{Lazarian}, A., \& {Xu}, S. 2021, \apj, 923, 53,
  \dodoi{10.3847/1538-4357/ac2de9}

\bibitem[{{Liu} {et~al.}(2022){Liu}, {Zeng}, {Xin}, \& {Zhang}}]{liu22}
{Liu}, S., {Zeng}, H., {Xin}, Y., \& {Zhang}, Y. 2022, Reviews of Modern Plasma
  Physics, 6, 19, \dodoi{10.1007/s41614-022-00080-6}

\bibitem[{{Ma} {et~al.}(2019){Ma}, {Wang}, {Zhang}, {Li}, \& {Yang}}]{ma19}
{Ma}, Y., {Wang}, H., {Zhang}, M., {Li}, C., \& {Yang}, J. 2019, \apj, 878, 44,
  \dodoi{10.3847/1538-4357/ab1ea7}

\bibitem[{{Malkov} \& {Aharonian}(2019)}]{Malkov_steep}
{Malkov}, M.~A., \& {Aharonian}, F.~A. 2019, \apj, 881, 2,
  \dodoi{10.3847/1538-4357/ab2c01}

\bibitem[{{Malkov} {et~al.}(2013){Malkov}, {Diamond}, {Sagdeev}, {Aharonian},
  \& {Moskalenko}}]{malkov13}
{Malkov}, M.~A., {Diamond}, P.~H., {Sagdeev}, R.~Z., {Aharonian}, F.~A., \&
  {Moskalenko}, I.~V. 2013, \apj, 768, 73, \dodoi{10.1088/0004-637X/768/1/73}

\bibitem[{{Malkov} \& {Drury}(2001)}]{MalkovDrury}
{Malkov}, M.~A., \& {Drury}, L.~O. 2001, Reports on Progress in Physics, 64,
  429, \dodoi{10.1088/0034-4885/64/4/201}

\bibitem[{{Reed} {et~al.}(1995){Reed}, {Hester}, {Fabian}, \&
  {Winkler}}]{reed95}
{Reed}, J.~E., {Hester}, J.~J., {Fabian}, A.~C., \& {Winkler}, P.~F. 1995,
  \apj, 440, 706, \dodoi{10.1086/175308}

\bibitem[{{Reynoso} \& {Goss}(2002)}]{casa_mc}
{Reynoso}, E.~M., \& {Goss}, W.~M. 2002, \apj, 575, 871, \dodoi{10.1086/341480}

\bibitem[{{Schure} \& {Bell}(2013)}]{schure13}
{Schure}, K.~M., \& {Bell}, A.~R. 2013, \mnras, 435, 1174,
  \dodoi{10.1093/mnras/stt1371}

\bibitem[{{Tuffs} {et~al.}(1997){Tuffs}, {Drury}, {Fischera}, {Heinrichsen},
  {Rasmussen}, {Russel}, \& {V{\"o}lk}}]{tuffs97}
{Tuffs}, R.~J., {Drury}, L.~O., {Fischera}, J., {et~al.} 1997, in ESA Special
  Publication, Vol. 419, The first ISO workshop on Analytical Spectroscopy, ed.
  A.~M. {Heras}, K.~{Leech}, N.~R. {Trams}, \& M.~{Perry}, 177

\bibitem[{{Weil} {et~al.}(2020){Weil}, {Fesen}, {Patnaude}, {Raymond},
  {Chevalier}, {Milisavljevic}, \& {Gerardy}}]{weil20}
{Weil}, K.~E., {Fesen}, R.~A., {Patnaude}, D.~J., {et~al.} 2020, \apj, 891,
  116, \dodoi{10.3847/1538-4357/ab76bf}

\bibitem[{{Xu} \& {Lazarian}(2022)}]{xu22}
{Xu}, S., \& {Lazarian}, A. 2022, \apj, 925, 48,
  \dodoi{10.3847/1538-4357/ac3824}

\bibitem[{{Yuan} {et~al.}(2017){Yuan}, {Lin}, {Fang}, \& {Bi}}]{yuan17}
{Yuan}, Q., {Lin}, S.-J., {Fang}, K., \& {Bi}, X.-J. 2017, \prd, 95, 083007,
  \dodoi{10.1103/PhysRevD.95.083007}

\bibitem[{{Yuan} {et~al.}(2013){Yuan}, {Funk}, {J{\'o}hannesson}, {Lande},
  {Tibaldo}, \& {Uchiyama}}]{yuan13}
{Yuan}, Y., {Funk}, S., {J{\'o}hannesson}, G., {et~al.} 2013, \apj, 779, 117,
  \dodoi{10.1088/0004-637X/779/2/117}

\bibitem[{{Zhou} {et~al.}(2018){Zhou}, {Li}, {Zhang}, {Vink}, {Chen}, {Arias},
  {Patnaude}, \& {Bregman}}]{zhou18}
{Zhou}, P., {Li}, J.-T., {Zhang}, Z.-Y., {et~al.} 2018, \apj, 865, 6,
  \dodoi{10.3847/1538-4357/aad960}

\bibitem[{{Zirakashvili} {et~al.}(2014){Zirakashvili}, {Aharonian}, {Yang},
  {O{\~n}a-Wilhelmi}, \& {Tuffs}}]{zirak14}
{Zirakashvili}, V.~N., {Aharonian}, F.~A., {Yang}, R., {O{\~n}a-Wilhelmi}, E.,
  \& {Tuffs}, R.~J. 2014, \apj, 785, 130, \dodoi{10.1088/0004-637X/785/2/130}

\end{thebibliography}
\bibliographystyle{aasjournal}
\end{document}